\documentclass[12pt]{article}
\usepackage{graphicx}
\usepackage[cp1251]{inputenc}
\usepackage{epsfig}
\usepackage[english]{babel}

\date{}

\begin{document}
\setcounter{page}{1}
\pagestyle{plain}

\title{\bf{Effects of doping and bias voltage on the screening in AAA-stacked trilayer graphene}}

\author{Yawar Mohammadi$^1$\thanks{Corresponding author. Tel./fax: +98 831 427
4569, Tel: +98 831 427 4569. E-mail address:
yawar.mohammadi@gmail.com} , Rostam Moradian$^{2}$, Farzad
Shirzadi Tabar$^{2}$} \maketitle{\centerline{$^1$Department of
Physics, Islamic Azad University, Kermanshah Branch, Kermanshah,
Iran} \maketitle{\centerline{$^2$Department of Physics, Razi
University, Kermanshah, Iran}

\begin{abstract}

We calculate the static polarization of AAA-stacked trilayer
graphene (TLG) and study its screening properties within the
random phase approximation (RPA) in all undoped, doped and biased
regimes. We find that the static polarization of undoped
AAA-stacked TLG is a combination of the doped and undoped
single-layer graphene static polarization. This leads to an
enhancement of the dielectric background constant along a
Thomas-Fermi screening with the Thomas-Fermi wave vector which is
independent of carrier concentrations and a $1/r^{3}$ power law
decay for the long-distance behavior of the screened coulomb
potential. We show that effects of a bias voltage can be taken
into account by a renormalization of the interlayer hopping energy
to a new bias-voltage-dependent value, indicating screening
properties of AAA-stacked TLG can be tuned electrically. We also
find that screening properties of doped AAA-stacked TLG, when
$\mu$ exceeds $\sqrt{2}\gamma$, are similar to that of doped SLG
only depending on doping. While for $\mu<\sqrt{2}\gamma$, its
screening properties are combination of SLG and AA-stacked bilayer
graphene screening properties and they are determined by doping
and the interlayer hopping energy.
\end{abstract}

%{\it \emph{PACS}}: \emph{74.20.-z, 74.20.Fg}

\vspace{0.5cm}

\section{Introduction}
\label{sec:1}

Single layer graphene (SLG), due to its linear low energy band
structure, shows many unusual properties which have not been
observed in the normal two dimensional electron gas
(2DEG)~\cite{b.1,b.2}. Furthermore, properties of multilayer
graphene materials, as have been found in the recent researches,
are dependent on their  stacking order and the number of
layers~\cite{b.3,b.4,b.5,b.6,b.7,b.8,b.9,b.10,b.11}. The most of
these researchs have been devoted to the few-layer graphene
materials with Bernal (ABA) and rhombohedral (ABC) stacking order.
Recently a new stable stacking order of few-layer graphene (with
AA stacking order) has been observed in experimental
researches~\cite{b.12,b.13}. In a few-layer graphene with AA
stacking order, each sublattice in a top layer is located directly
above the same one in the bottom layer. These materials, due to
this staking order, have a special low energy band structure which
is a composition of electron-doped, hole-doped and even undoped
SLG-like band structures~\cite{b.13,b.14}. For some properties,
these SLG-like bands behave like decoupled bands leading to many
attractive properties which are different form that of SLG and
also not been observed in the other graphene-based
materials~\cite{b.14,b.15,b.16,b.17,b.18,b.19,b.20}.

One of the most attractive physical quantities of a system is the
static polarization. Obtaining the static polarization is
essential to study many fundamental properties, e.g., the screened
coulomb interaction of charged impurities, the RKKY interaction
between magnetic adatoms and Kohn anomaly in phonon dispersion.
Recently several groups have studied the static polarization of
SLG~\cite{b.21,b.22,b.23,b.24,b.25} and bilayer graphene (BLG)
with both AB~\cite{b.26,b.27,b.28} and AA~\cite{b.29} stacking
order and the other few-layer-graphene
materials~\cite{b.7,b.30,b.31,b.32,b.33,b.34}. These works showed
that the static polarization and the screening effects in these
materials not only are dependent on the number of layers and the
stacking order but also have unusual futures not been observed so
far. Motivated by these facts, we calculate the static
polarization of AAA-stacked TLG and study its screening properties
in this paper.

The rest of this paper is organized as follows. In the section II,
first we introduce the tight-binding Hamiltonian describing the
low energy quasiparticle excitation in AAA-stacked TLG. Then we
obtain a relation for the dynamical polarization which can be
solved analytically to obtain the static polarization. In the next
section we present our results for the static dielectric function
and the screened coulomb potential in several cases. First we
study them in undoped AAA-stacked TLG. Then we discuss how one can
apply a bias voltage to the system and consider effects of the
bias voltage on the screening properties of AAA-stacked TLG. At
the end of this section we investigate effects of doping. Finally,
we end our paper by summary and conclusions at section IV.

\section{Model Hamiltonian}
\label{sec:2}

In an AAA-stacked TLG lattice which is composed of three SLG, each
sublattice in a top layer is located directly above the same one
in the bottom layer. The unit cell of an AAA-stacked TLG consists
of 6 inequivalent Carbon atoms, two atoms for every layer. Thus
its Hamiltonian, in the nearest-neighbor tight-binding
approximation, is given by
\begin{eqnarray}
H=\sum_{\mathbf{q}}\hat{\Psi}^{\dag}_{\mathbf{q}}\widehat{H}_{\mathbf{q}}
\hat{\Psi}_{\mathbf{q}}, \label{eq:01}
\end{eqnarray}
where
\begin{eqnarray}
\widehat{H}_{\mathbf{q}} = \left(
\begin{array}{cccccc}
0 &   \phi^{\ast}(\mathbf{q})   & \gamma & 0 & 0 & 0 \\
\phi(\mathbf{q})   & 0 & 0 & \gamma & 0 & 0 \\
\gamma & 0 & 0 & \phi^{\ast}(\mathbf{q}) & \gamma & 0 \\
0 & \gamma & \phi(\mathbf{q}) & 0 & 0 & \gamma \\
0 & 0 & \gamma & 0 & 0 &    \phi^{\ast}(\mathbf{q}) \\
0 & 0 & 0 & \gamma &   \phi(\mathbf{q})   & 0 \\
\end{array}
\right),\label{eq:02}
\end{eqnarray}
and
$\hat{\Psi}^{\dag}_{\mathbf{q}}=(a^{\dag}_{1\mathbf{q}},b^{\dag}_{1\mathbf{q}}
,a^{\dag}_{2\mathbf{q}},b^{\dag}_{2\mathbf{q}},a^{\dag}_{3\mathbf{q}}
,b^{\dag}_{3\mathbf{q}})$. Here $a^{\dag}_{n\mathbf{q}}$
($b^{\dag}_{n\mathbf{q}}$) create an electron with momentum
$\mathbf{q}$ at A(B) sublattice in nth-layer. $\gamma$ is the
interlayer hopping energy and
$\phi(\mathbf{q})=-t\sum_{i=1}^{^{3}}e^{i\mathbf{q}.\mathbf{d}_{i}}$
where $\mathbf{q}=(q_{x},q_{y})$ is two dimensional momentum and
$\mathbf{d}_{1}=(a\sqrt{3}/2,a/2)$,
$\mathbf{d}_{2}=(-a\sqrt{3}/2,a/2)$ and $\mathbf{d}_{3}=(0,-a)$
are vectors connect each Carbon atom to its in-plane nearest
neighbors with $a$ being the shortest Carbon-Carbon distance.
AAA-stacked TLG, similar to SLG, has a two dimensional hexagonal
Brilloin zone and its low-energy excitations occur near Dirac
points ($\mathbf{K}$ and $\mathbf{K}^{'}$). To obtain the
Hamiltonian dominating low-energy excitations, we must expand
$\phi(\mathbf{q})(\phi^{\ast}(\mathbf{q}))$ around Dirac points
for $|\mathbf{q} |\ll|\mathbf{K}|$ (where
$\mathbf{q}=\mathbf{k}+\mathbf{K}$). By expanding them around
$\mathbf{K}$ point, we have
$\phi(\mathbf{q})=v_{F}k_{+}(\phi^{\ast}(\mathbf{q})=v_{F}k_{-})$
where $k_{\pm}=k_{x}\pm ik_{y}$ and $v_{F}=3ta/2$ is Fermi
velocity. Now we apply the unitary transformation
$\widehat{U}^{-1}\widehat{H}_{\mathbf{k}}\widehat{U}$, where
\begin{eqnarray}
\widehat{U} = \left(
\begin{array}{cccccc}
\sqrt{2} &   0   & 2 & 0 & \sqrt{2} & 0 \\
0   & \sqrt{2} & 0 & 2 & 0 & \sqrt{2} \\
2 & 0 & 0 & 0 & -2 & 0 \\
0 & 2 & 0 & 0 & 0 & -2 \\
\sqrt{2} & 0 & -2 & 0 & \sqrt{2} &  0 \\
0 & \sqrt{2} & 0 & -2 & 0 & \sqrt{2} \\
\end{array}
\right),\label{eq:03}
\end{eqnarray}
to rewrite the low-energy Hamiltonian in a block-diagonal form as
\begin{eqnarray}
\widehat{H}_{\mathbf{k}} = \left(
\begin{array}{cccccc}
\sqrt{2}\gamma & v_{F}k_{-} & 0 & 0 & 0 & 0 \\
v_{F}k_{+} & \sqrt{2}\gamma & 0 & 0 & 0 & 0 \\
0 & 0 & 0 & v_{F}k_{-} & 0 & 0 \\
0 & 0 & v_{F}k_{+} & 0 & 0 & 0 \\
0 & 0 & 0 & 0 & -\sqrt{2}\gamma & v_{F}k_{-} \\
0 & 0 & 0 & 0 & v_{F}k_{+} & -\sqrt{2}\gamma \\
\end{array}
\right),\label{eq:04}
\end{eqnarray}
and to obtain easily whose low-energy eigenvalues and eigenstates,
\begin{eqnarray}
\varepsilon_{\mathbf{k}}^{0\lambda}=\lambda v_{F}|\mathbf{k}| &,&
\phi^{0\lambda}_{\mathbf{k}}=\frac{1}{\sqrt{2}} \left(
\begin{array}{c}
0\\
0\\
1\\
\lambda e^{-i\theta_{\mathbf{k}}}\\
0\\
0\\
\end{array}
\right),\label{eq:05}
\end{eqnarray}
\begin{eqnarray}
\varepsilon_{\mathbf{k}}^{+\lambda}=+\sqrt{2}\gamma+\lambda
v_{F}|\mathbf{k}| &,&
\phi^{+\lambda}_{\mathbf{k}}=\frac{1}{\sqrt{2}} \left(
\begin{array}{c}
1\\
\lambda e^{-i\theta_{\mathbf{k}}}\\
0\\
0\\
0\\
0\\
\end{array}
\right),\label{eq:06}
\end{eqnarray}
and
\begin{eqnarray}
\varepsilon_{\mathbf{k}}^{-\lambda}=-\sqrt{2}\gamma+\lambda
v_{F}|\mathbf{k}| &,&
\phi^{-\lambda}_{\mathbf{k}}=\frac{1}{\sqrt{2}} \left(
\begin{array}{c}
0\\
0\\
0\\
0\\
1\\
\lambda e^{-i\theta_{\mathbf{k}}}\\
\end{array}
\right),\label{eq:07}
\end{eqnarray}
where $\lambda=\pm$, $|\mathbf{k}|=\sqrt{k_{x}^{2}+k_{y}^{2}}$ and
$\theta_{\mathbf{k}}=tan^{-1}(k_{y}/k_{x})$. The low energy
density of states of AAA-stacked TLG is
\begin{eqnarray}
D(E)=g\frac{|E-\sqrt{2}\gamma|+2|E|+|E+\sqrt{2}\gamma|}{8\pi
v_{F}^{2}},\label{eq:08}
\end{eqnarray}
where multiple $g=4$ is due to spin and valley degeneracies. The
low energy band structure and the density of states of AAA-stacked
TLG have been shown in Fig. \ref{fig:01}.

To consider screening properties of AAA-stacked TLG, we must
calculate the static polarization. The total static polarization
is the static limit of the total dynamic polarization,
$\Pi(\mathbf{\mathbf{q}},\omega\rightarrow0)$. The equivalent
Matsubara function for the total dynamic polarization is given by
\begin{eqnarray}
\Pi(\mathbf{q},i\omega_{n})=-\frac{1}{A}\int_{0}^{1/T} d\tau
e^{i\omega_{n}\tau} \langle T_{\tau}
\rho(\mathbf{q},\tau)\rho(-\mathbf{q},0)\rangle,\label{eq:09}
\end{eqnarray}
where $A$ is lattice area, $\omega_{n}$ are bosonic Matsubara
frequencies and $T$ is the temperature. The retarded function for
the static polarization is obtained from the Matsubara function by
changing $i\omega_{n}$ to $\omega+i0^{+}$. By making use of the
definition of the Green's function, Eq. (\ref{eq:09}) becomes
\begin{eqnarray}
\Pi(\mathbf{q},i\omega_{n})=-\frac{T}{A}\sum_{m \mathbf{k}}
Tr[\hat{G}(\mathbf{k+q},i\nu_{m}+i\omega_{n})\hat{G}(\mathbf{k},i\nu_{m})].\label{eq:10}
\end{eqnarray}
We use the spectral function representation of the Matsubara
Green's function,
\begin{eqnarray}
\hat{G}(\mathbf{k},i\omega_{n})=\int_{-\infty}^{+\infty}\frac{d\Omega}{2\pi}
\frac{\hat{A}(\mathbf{k},\Omega)}{i\omega_{n}-\Omega},\label{eq:11}
\end{eqnarray}
and apply the Matsubara summation to obtain the following relation
for the dynamic polarization
\begin{eqnarray}
\Pi(\mathbf{q},i\omega_{n})=-\frac{g}{4A}[2\sum_{\mathbf{k},\lambda\lambda^{'}
}\frac{n_{\mathbf{k}}^{0\lambda}-n_{\mathbf{k+q}}^{0\lambda^{'}}}{
i\omega_{n}+\varepsilon_{\mathbf{k}}^{0\lambda}-\varepsilon_{\mathbf{k+q}}^{0\lambda^{'}}}
+ \sum_{\mathbf{k},\lambda\lambda^{'}s=\pm
}\frac{n_{\mathbf{k}}^{s\lambda}-n_{\mathbf{k+q}}^{s\lambda^{'}}}{
i\omega_{n}+\varepsilon_{\mathbf{k}}^{s\lambda}-\varepsilon_{\mathbf{k+q}}^{s\lambda^{'}}}]
\times[1+\lambda\lambda^{'}\cos
\Delta\theta_{\mathbf{k}+\mathbf{q},\mathbf{k}}],\label{eq:12}
\end{eqnarray}
where
$n_{\mathbf{k}}^{s\lambda}=1/(1+exp[(\varepsilon_{\mathbf{k}}^{s\lambda}-\mu)/T])$
is the Fermi-Dirac distribution function and
$\Delta\theta_{\mathbf{k}+\mathbf{q},\mathbf{k}}=\theta_{\mathbf{k}+\mathbf{q}}-\theta_{\mathbf{k}}$.

One of the most attractive quantities, which can be calculated
easily from the static polarization, is the screening of a charged
impurity $Ze$. If the impurity is located at the center of the
coordinate set, the intra-layer screened coulomb potential at
distance $r$ from the impurity is $\phi(r)=\int qdq \phi(q)
J_{0}(qr)/2\pi$, where $J_{0}(qr)$ is the Bessel function of the
zeroth order and $\phi(q)$ is the Fourier transform of the
screened coulomb potential. $\phi(q)$, at low energy excitations
and within RPA, is given by $\phi(q)\approx2\pi Ze^{2}/(\kappa
q[1+(2\pi e^{2}/\kappa q)\Pi(q)])$ where $\Pi(q)$ is the total
static polarization. In the next section we present our results
for the static polarization and consider screening properties of
AAA-stacked TLG.

\section{Numerical Results}
\label{sec:4}

In this section, we consider screening properties of AAA-stacked
TLG in several cases, in both undoped and doped cases and also in
the presence of a bias voltage. In each case, first we calculate
the static polarization. Then we study the Thomas-Fermi screening
and the long-distance behavior of the screened coulomb potential.
Our results are presented in the following subsections.

\subsection{Undoped AAA-stacked TLG}

In AAA-stacked TLG, similar to AA-stacked BLG and as Eq.
(\ref{eq:12}) shows, only those transition contribute to the total
static polarization that occur between bands with same s index.
This is similar to what has been reported for the optical
conductivity of AA-stacked BLG~\cite{b.17}, where energy bands
with different s index behave like decoupled bands. It is evident
from Eq. (\ref{eq:12}) that the static polarization of AA-stacked
TLG can be written as sum of four SLG-like terms for the static
polarization, two terms with $\mu=0$ and two other terms with
$\mu=\sqrt{2}\gamma$ and $\mu=-\sqrt{2}\gamma$ (all of them are
weighted by $1/4$). Due to equality of later terms, we can write
$\Pi(q)=\Pi_{SLG}^{\mu=0}(q)/2+
\Pi_{SLG}^{\mu=\sqrt{2}\gamma}(q)/2$. At zero temperature,
$\Pi(q)$ is given by (Appendix A)
\begin{eqnarray}
\Pi(q)=\frac{1}{2}\frac{g q}{16 v_{F}}+\frac{1}{2}\frac{\sqrt{2}
g\gamma}{2\pi v_{F}^{2}}
 ,\label{eq:13}
\end{eqnarray}
for $q\leq 2\sqrt{2}\gamma/v_{F}$, and
\begin{eqnarray}
\Pi(q)=\frac{1}{2}\frac{g q}{16 v_{F}}+\frac{1}{2}[\frac{g q}{16
v_{F}}+\frac{\sqrt{2}g\gamma}{2\pi
v_{F}^{2}}(1-\frac{1}{2}\sqrt{1-(\frac{2\sqrt{2}\gamma}{v_{F}q})^{2}}
-\frac{v_{F}q}{4\sqrt{2}\gamma}\sin^{-1}\frac{2\sqrt{2}\gamma}{v_{F}q})]
 ,\label{eq:14}
\end{eqnarray}
for $q> 2\sqrt{2}\gamma/v_{F}$. It is evident that, for
$q\leq\sqrt{2}\gamma$, the static polarization is a combination of
a constant metallic-like polarization and an insulating-like
polarization increasing linearly with $q$. This is a very
different behavior from the ordinary two dimensional electron gas
(2DEG)~\cite{b.35} and from other graphene-based
materials~\cite{b.21,b.22,b.23,b.24,b.25,b.26,b.27,b.28,b.29,b.30,b.31,b.32,b.33,b.34}.
While for $q>\sqrt{2}\gamma$, that the interband transitions
between bands with same s index dominate the intraband
transitions, the static polarization of AAA-stacked TLG (similar
to that in SLG~\cite{b.21,b.23,b.25}) increases linearly with $q$.

The static dielectric function is written as $\epsilon(q)=1+(2\pi
e^{2}/\kappa q)\Pi(q)$ where $\kappa$ is the background dielectric
constant. The linear part of the static polarization only leads to
an enhancement of the effective background dielectric constant
$\kappa^{\ast}=\kappa(1+\pi e^{2}g/16\kappa v_{F})$. While its
constant metallic-like part (in the long-wavelength limit) gives
rise to the Thomas-Fermi screening with $\epsilon(q)=1+q_{TF}/q$
where $q_{TF}=2\pi e^{2} D(0)/\kappa^{\ast}=\sqrt{2}g\gamma
e^{2}/2 \kappa^{\ast} v_{F}^{2}$ is the effective Thomas-Fermi
screening wave vector. Note that the effective Thomas-Fermi
screening wave vector in undoped AAA-stacked TLG, similar to that
in the ordinary 2DEG~\cite{b.35} and AA-stacked BLG~\cite{b.29}
(and contrary to that in SLG~\cite{b.21,b.25}), is independent of
the carrier concentration.

Another of interest quantity is the long-distance behavior of the
screened coulomb potential from a charged impurity. This is
determined by the long-wavelength behavior of the static
polarization and its singularity. The long-wavelength behavior of
the static polarization leads to a non-oscillatory term for the
screened coulomb potential as~\cite{b.22,b.26},
\begin{eqnarray}
\phi(r)=\frac{Ze^{2}}{\kappa^{\ast}r}-\frac{\pi Ze^{2}q_{TF}
}{2\kappa^{\ast}}[H_{0}(q_{TF}r)-Y_{0}(q_{TF}r)],\label{eq:15}
\end{eqnarray}
where $H_{0}$ and $Y_{0}$ are the Struve and the Bessel functions
of the second kind and its asymptotic form at large distance is
given by $Ze^{2}q_{TF}/[\kappa^{\ast}(q_{TF}r)^{3}]$. Furthermore,
there is a discontinuity occurring in the second derivative of the
static polarization ( as q approaches $2\sqrt{2}\gamma/v_{F}$ from
above $d^{2}\Pi(q)/d^{2}q~\propto~1/\sqrt{1-(2\sqrt{2}\gamma/q
v_{F})^{2}}$), which leads to a Friedel-oscillation term for the
screened coulomb potential. This term can be evaluated using a
theorem of Lighthill~\cite{b.36}, and whose asymptotic form is $
k_{F}^{'}cos(2k_{F}^{'}r)/\kappa^{\ast}(k_{F}^{'}r)^{3}$ where
$k_{F}^{'}=\sqrt{2}\gamma/v_{F}$(Appendix B). For AAA-stacked TLG,
both terms are of same order in distance and must be kept to
discuss the long-distance of the screened coulomb potential.

\subsection{Effects of a bias voltage}

Effects of a bias voltage can be considered via adding an electric
potential matrix,
\begin{eqnarray}
\widehat{V}=V\left(
\begin{array}{ccc}
\hat{1} &  \hat{0} & \hat{0} \\
\hat{0}  & \hat{0} & \hat{0} \\
\hat{0} & \hat{0} & -\hat{1} \\
\end{array}
\right),\label{eq:16}
\end{eqnarray}
to the Hamiltonian matrix in Eq. (\ref{eq:02}), where $\hat{1}$
and $\hat{0}$ are $2\times2$ unitary and zero matrix respectively.
$2V$ is the difference between the applied electrical potential to
the top and bottom layer. This leads to a new dispersion relation
for AAA-stacked TLG as
\begin{eqnarray}
\varepsilon_{V,\mathbf{k}}^{s\lambda}=s
[(\sqrt{2}\gamma)^{2}+V^{2}]^{1/2}+\lambda
v_{F}|\mathbf{k}|,\label{eq:17}
\end{eqnarray}
which indicates that effects of the bias voltage can be taken into
account by a renormalization of the interlayer hopping energy of
the unbiased AAA-stacked TLG, $\gamma$, to a new
bias-voltage-dependent hopping energy,
$\gamma^{'}=[\gamma^{2}+V^{2}/2]^{1/2}$. Moreover, it is easy to
show that Eq. (\ref{eq:01}) is also indefeasible for the dynamical
polarization of the biased AAA-stacked TLG, but by replacing
$\varepsilon_{\mathbf{k}}^{s\lambda}\rightarrow\varepsilon_{V,\mathbf{k}}^{s\lambda}$.

The static polarization of biased AAA-stacked TLG is obtained from
Eqs. (\ref{eq:13}) and (\ref{eq:14}), by just substituting
$\gamma\longleftrightarrow \gamma^{'}$. Thus it, for
$q<\sqrt{2}\gamma^{'}/v_{F}$, is a combination of a constant
metallic-like polarization and an insulating-like polarization
increasing linearly with $q$. But it increases linearly with $q$
when $q<\sqrt{2}\gamma^{'}/v_{F}$. The linear part of the static
polarization (similar to the undoped case) gives rise to an
enhancement of the effective background dielectric constant
$\kappa^{\ast}=\kappa(1+\pi e^{2}g/16\kappa v_{F})$ which is not
affected by the bias voltage. In the biased AAA-stacked TLG, the
effective Thomas-Fermi screening wave vector depends on the
applied bias voltage as
$q_{TF}=\sqrt{2}ge^{2}[\gamma^{2}+V^{2}/2]^{1/2}/2 \kappa^{\ast}
v_{F}^{2}$, which indicates enhanced screening at high bias
voltage. This also leads to a fast decayed screened coulomb
potential as $\phi(r)\propto 1/[\gamma^{2}+V^{2}/2]$ at high bias
voltage.

\subsection{Effects of doping}

Considering effects of doping on screening properties of
AAA-stacked TLG is more attractive than the other cases. It is
evident from Eq. (\ref{eq:01}) that, for $\mu>\sqrt{2}\gamma$, the
static polarization of AAA-stacked is equal to that of a doped SLG
lattice, a constant metallic polarization for $q\leq\mu/v_{F}$ and
a linear in-q insulating polarization for $q>\mu/v_{F}$. Thus,
when $\mu$ exceeds $\sqrt{2}\gamma$, screening properties of
AAA-stacked TLG is similar to that of doped
SLG~\cite{b.21,b.22,b.23,b.24,b.25,b.26}. But when $\mu$ is less
than $\sqrt{2}\gamma$, the static polarization of AAA-stacked TLG
is given by $\Pi(q)=\Pi_{SLG}^{\mu}(q)/2+
\Pi_{SLG}^{\sqrt{2}\gamma}(q)/2$ which is a constant metallic-like
polarization (for $q\leq\mu/v_{F}$) and a linear in-q
insulating-like one (when $q$ exceeds $\sqrt{2}\gamma/v_{F}$),
while, for $\mu/v_{F}<q\leq\sqrt{2}\gamma/v_{F}$, it is a
combination of the constant metallic-like and the linear in-q
insulating-like polarization. This is in contrast to what has been
reported for the ordinary 2DEG~\cite{b.21} and for the other
graphene-based
materials~\cite{b.21,b.22,b.23,b.24,b.25,b.26,b.27,b.33}.

The long-wavelength limit of the static dielectric function is
given by $\epsilon(q)\approx 1+q_{TF}/q$, where $q_{TF}=2\pi
e^{2}D(0)/\kappa=ge^{2}(\mu+\sqrt{2}\gamma)/2\kappa v_{F}^{2}$.
Note that the Thomas-Fermi screening wave vector depends on doping
and the interlayer hopping energy. This indicates that screening
properties of AAA-stacked TLG neither are only determined by
doping (similar to that in SLG) nor are only determined by the
interlayer hopping energy (similar to that in AA-stacked BLG). It
is evident that, in addition to doping (in SLG ) and the
interlayer hopping energy (in AA-stacked BLG), screening
properties also depends on the Fermi velocity, $v_{F}$.

In the doped AAA-stacked TLG, the screened Coulomb potential has
another Freidel-oscillation term coming from the discontinuity of
$d^{2}\Pi(q)/d^{2}q$ at $q=k_{F}=\mu/v_{F}$. This term at large
distance decays as $k_{F}\cos(2k_{F}r)/\kappa(k_{F}r)^{3}$ and it
can be tuned by doping. While the other Friedel-oscillation term,
$k_{F}^{'}\cos(2k_{F}^{'})/\kappa(k_{F}^{'}r)^{3}$, only depends
on the interlayer hopping energy and it is not affected by doping.
The non-oscillatory part of the screened coulomb potential which
decays at large distances as $eq_{TF}/[\kappa(q_{TF}r)^{3}]$, due
to partial dependence of $eq_{TF}$ on $\mu$, can be suppressed by
doping partially.

\section{Summary and conclusions}
\label{sec:5}

In summary, we studied screening properties of a AAA-stacked TLG
lattice within RPA in all undoped, doped and biased regimes. We
found that in undoped case, the static polarization of AAA-stacked
TLG is a combination of a constant metallic-like and an
insulating-like polarization, leading to an enhancement of the
effective background dielectric constant along a Thomas-Fermi
screening with the Thomas-Fermi wave vector which, similar to that
in ordinary 2DEG and undoped AA-stacked BLG, is independent of the
carrier concentrations. We also discussed the long-distance
behavior of the screened Coulomb potential determined by two
terms, a non-oscillatory term coming from the long-wavelength
limit of the static polarization and a Freidel-oscillation term
coming from the discontinuity in the second derivative of the
polarization, whose asymptotic forms at large distances are
$1/r^{3}$ and $cos(2k_{F}^{'}r)/r^{3}$ respectively.

Moreover, we studied effects of a bias voltage and showed that it
can be taken into account by a renormalization of the interlayer
hopping energy to a new value, $\gamma \rightarrow
\sqrt{\gamma^{2}+V^{2}/2}$. We showed that screening properties of
biased AAA-stacked TLG, e.g., the Thomas-Fermi wave vector and the
long-distance behavior of the screened coulomb potential, can be
tuned by the bias voltage.

At the final subsection, we considered screening properties of
doped AAA-stacked TLG. We showed that when $\mu$ exceeds
$\sqrt{2}\gamma$, screening properties of doped AAA-stacked TLG is
similar to that of doped SLG. But for doped AAA-satcked TLG with
$\mu\leq\sqrt{2}\gamma$, screening properties are dependent on the
inerlayer hopping energy and doping, in contrast to that in SLG
only depending on doping and to that in AA-stacked BLG only
depending on the inerlayer hopping energy.

\section{Appendix A: Calculating $\Pi(q,0)$}

In the undoped case, $\Pi(q,\omega)$ can be written as
\begin{eqnarray}
\Pi(q,\omega)=\frac{1}{4}(2\Pi^{ud}(q,\omega)+\Pi^{pd}(q,\omega)+\Pi^{hd}(q,\omega))\nonumber
.~~~~~~~~~~~~~~~~~~~~~~~~~~~~~~~(A.1)\label{eq:A1}
\end{eqnarray}

The imaginary part of $\Pi^{ud}(q,\omega)$ is given by
\begin{eqnarray}
Im\Pi^{ud}(q,\omega)=g\int_{0}^{\infty}\frac{k
dk}{8\pi}\int_{0}^{2\pi} d\varphi
\frac{|\mathbf{k+q}|-k-q\cos\varphi}{|\mathbf{k+q}|}\sum_{\lambda=\pm}
\lambda \delta(\omega+\lambda v_{F}(k+|\mathbf{k+q}|))
\nonumber.~~~~(A.2)\label{eq:A2}
\end{eqnarray}

The $\varphi$-integration yields
\begin{eqnarray}
Im\Pi^{ud}(q,\omega)=g\int_{\frac{\frac{\omega}{v_{F}}-q}{2}}^{\frac{\frac{\omega}{v_{F}}+q}{2}
}\frac{dk}{4\pi}
\sqrt{\frac{(\frac{\omega}{v_{F}}-2k)^{2}-q^{2}}{(\frac{\omega}{v_{F}})^{2}-q^{2}}}
\nonumber ,~~~~~~~~~~~~~~~~~~~~~~~(A.3) \label{eq:A3}
\end{eqnarray}
which after integration over $k$ becomes
\begin{eqnarray}
Im\Pi^{ud}(q,\omega)=\frac{g
q^{2}}{16v_{F}\sqrt{(\omega/v_{F})^{2}-q^{2}}}\nonumber.~~~~~~~~~~~~~~~~~~~~~~~~(A.4)
\label{eq:A4}
\end{eqnarray}

Using the Kramers-Kornig relation, the real part of
$\Pi^{ud}(q,\omega)$,
\begin{eqnarray}
Re\Pi^{ud}(q,\omega)=\frac{g
q^{2}}{16v_{F}\sqrt{q^{2}-(\omega/v_{F})^{2}}}\nonumber
,~~~~~~~~~~~~~~~~~~~~~~~~(A.5) \label{eq:A5}
\end{eqnarray}
is obtained, indicating
$\Pi^{ud}(q)=\Pi_{SLG}^{\mu=0}(q)=\frac{gq}{16v_{F}}$~\cite{b.21,b.22,b.25}.
The particle-doped part of $\Pi(q,\omega)$ can be written as
\begin{eqnarray}
\Pi^{pd}(q)=\frac{gq}{16v_{F}}+ g\int_{0}^{\infty}
\frac{dk}{2\pi^{2}v_{F}}
\theta(\sqrt{2}\gamma-v_{F}k)[\int_{0}^{\pi} d\varphi (1-
\frac{1-(2k/q)^{2}} {1+2(k/q)\cos\varphi}] \nonumber.
~~~~~~~~~~~~~~~~~(A.6) \label{eq:A6}
\end{eqnarray}

The $\varphi$-integration yields
\begin{eqnarray}
\Pi^{pd}(q)=&& \frac{gq}{16v_{F}}+\frac{g\sqrt{2}\gamma}{2\pi
v_{F}^{2}} - \frac{g}{2\pi^{2}v_{F}} [ \int_{0}^{q/2}dk
(\theta(\sqrt{2}\gamma-v_{F}k)
[2uv\tan^{-1}(\frac{u}{v}\tan\frac{\varphi}{2}) |_{\varphi=0}^{\varphi=\pi}]) +\nonumber \\
&& \int_{q/2}^{\infty}dk
(\theta(\sqrt{2}\gamma-v_{F}k)[2uw\ln|\frac{u\cos\frac{\varphi}{2}
+w\sin\frac{\varphi}{2}}{u\cos\frac{\varphi}{2}-w\sin\frac{\varphi}{2}}
|_{\varphi=0}^{\varphi=\pi}])] \nonumber
,~~~~~~~~~~~~~~~~~~~~~~~~(A.7) \label{eq:A7}
\end{eqnarray}
where $u=\sqrt{1+2k/q}$ and $v=\sqrt{1-2k/q}=iw$. After
integrating over $k$ we have
\begin{eqnarray}
\Pi^{pd}(q)=\frac{gq}{16v_{F}}+\frac{g\sqrt{2}\gamma}{2\pi
v_{F}^{2}}-\frac{gq}{16 v_{F}} \nonumber,~~~~~~~~~~~~~~~~~~~~~~~~
(A.8) \label{eq:A8}
\end{eqnarray}
for $q<2\sqrt{2}\gamma/v_{F}$, and
\begin{eqnarray}
\Pi^{pd}(q)=\frac{gq}{16v_{F}}+\frac{g\sqrt{2}\gamma}{2\pi
v_{F}^{2}} - \frac{1}{2}\frac{g\sqrt{2}\gamma}{2\pi v_{F}^{2}} [
\sqrt{1-(\frac{2\sqrt{2}\gamma}{v_{F}q})^{2}}
+\frac{v_{F}q}{2\sqrt{2}\gamma}\sin^{-1}(\frac{2\sqrt{2}\gamma}{v_{F}q})]
\nonumber,~~~~~~~~~(A.9) \label{eq:A9}
\end{eqnarray}
for $q>2\sqrt{2}\gamma/v_{F}$. This is equal to
$\Pi_{SLG}^{\mu=\sqrt{2}\gamma}(q)$~\cite{b.25,b.25}. The
hole-doped part of $\Pi(q,\omega)$ is
\begin{eqnarray}
\Pi^{hd}(q)=\frac{1}{A}\sum_{\mathbf{k}}[ \frac{1+\cos
\Delta\theta_{\mathbf{k}+\mathbf{q},\mathbf{k}}}{
v_{F}(|\mathbf{k}|-|\mathbf{k+q}|)}
(n_{\mathbf{k}}^{+-}-n_{\mathbf{k+q}}^{+-}) + \frac{1-\cos
\Delta\theta_{\mathbf{k}+\mathbf{q},\mathbf{k}}}{
v_{F}(|\mathbf{k}|+|\mathbf{k+q}|)}(n_{\mathbf{k}}^{+-}+n_{\mathbf{k+q}}^{+-})]\nonumber,~~~(A.10)
\label{eq:A10}
\end{eqnarray}
where
$n_{\mathbf{k}}^{+-}=1-\theta(\sqrt{2}\gamma-v_{F}|\mathbf{k}|)$.
After substituting these terms into eq. (A.10), we get
\begin{eqnarray}
\Pi^{hd}(q)=\Pi^{ud}(q)- \frac{g}{A}\sum_{\mathbf{k}
}\theta(\sqrt{2}\gamma-v_{F}|\mathbf{k}|)[ \sum_{\lambda=\pm}
\frac{1-\lambda \cos
\Delta\theta_{\mathbf{k}+\mathbf{q},\mathbf{k}}}
{v_{F}(|\mathbf{k}|+\lambda|\mathbf{k+q}|)}] \nonumber
,~~~~~~~~~~~~~~~(A.11) \label{eq:A11}
\end{eqnarray}
which is equal to $\Pi^{pd}(q)$. The static polarization of biased
AAA-stacked TLG can be obtained easily as it has been calculated
in the undoped case just by substituting
$\gamma\longleftrightarrow \gamma^{'}$. Also the static
polarization of doped AAA-stacked TLG is calculated similar to
$\Pi^{pd}(q)$(or $\Pi^{hd}(q)$) of the static polarization in the
undoped case.

\section{Appendix B: Friedel oscillations} The Friedel
oscillation, which comes from singularities of the static
dielectric function, can be calculated by making use of a theorem
of Lighthill~\cite{b.36}. According to this theorem, the
asymptotic expression for the Fourier transform of function
$f(x)$, which has a finite number of singularities $x=x_{1},x_{2},
... ,x_{M} $, is given by
\begin{eqnarray}
g(y)=\sum_{m=1}^{M}G_{m}(y)+o(|y|^{-N})~~as~~|y|\rightarrow \infty
 \nonumber ,~~~~~~~~~~~~~~~~~~~~~~~(B.1)
\label{eq:B1}
\end{eqnarray}
where $g(y)$ is the Fourier transform of $f(x)$ and $G_{m}(y)$ are
the Fourier transform of $F_{m}(x)$ which make Nth derivative of
$f(x)-F_{m}(x)$ absolutely integrable in an interval including
$x_{m}$.

To obtain the asymptotic expression of the Friedel oscillation we
must use the asymptotic form of the zeroth-order of the Bessel
function. Therefore, we have
\begin{eqnarray}
\phi(r)\simeq Ze^{2} \frac{\sqrt{k_{F}}}{\sqrt{\pi r}}
\int_{0}^{\infty} \frac{\cos(k_{F}rx)+\sin(k_{F}rx)}{x+\frac{2\pi
e^{2}}{k_{F}}\Pi(x)}\sqrt{x}dx, \nonumber ~~~~~~~~~~~~~~(B.2)
\label{eq:B2}
\end{eqnarray}
where $k_{F}=\sqrt{2}\gamma/v_{F}$, $x=q/k_{F}$.

By making use of the theorem and the asymptotic behavior of the
Fourier transform of the function $\theta(x-2)\sqrt{|x-2|^{3}}$
which is
\begin{eqnarray}
g(y)\simeq \frac{3\sqrt{\pi}}{4(2\pi|y|)^{5/2}} e^{-i(2y+5\pi/4)}
\nonumber ~,~~~~~~~~~~~~~~~~~~~~~~~~~~~~(B.3) \label{eq:B3}
\end{eqnarray}
we obtain the asymptotic form of the Friedel oscillation,
\begin{eqnarray}
\phi(r)\simeq -\frac{3Ze^{2}}{4\kappa^{\ast}}
\frac{k_{F}^{'}\alpha}{(1+\pi \alpha)^{2}}
\frac{\cos(2k_{F}^{'}r)}{(k_{F}^{'}r)^{3}} \nonumber
~,~~~~~~~~~~~~~~~~~~~~~~~~~(B.4) \label{eq:B4}
\end{eqnarray}
where $k_{F}^{'}=\sqrt{2}\gamma/v_{F}$ and
$\alpha=e^{2}/\kappa^{\ast}v_{F}$. The Friedel oscillation coming
from the discontinuity $d^{2}\Pi(q)/d^{2}q$ at
$\sqrt{2}[\gamma^{2}+V^{2}/2]^{1/2}/v_{F}$(in the biased case) and
$q=\mu/v_{F}$(in the doped case) are obtained completely similar
to that in the undoped case.

%\newpage
%

%
%
\newpage
\begin{figure}
\begin{center}
\includegraphics[width=12cm,angle=0]{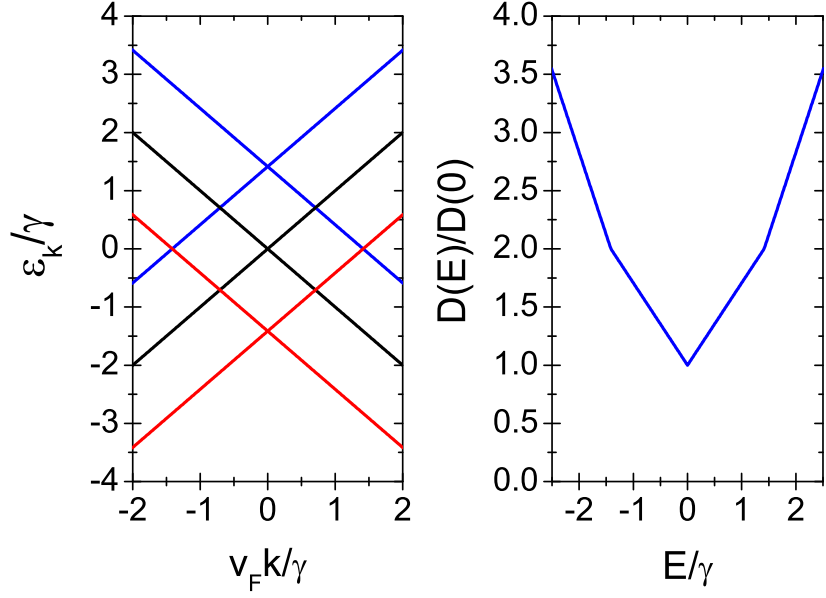}
\caption{Left panel shows the low energy band structure of
AAA-stacked TLG. Blue, black and red lines are correspond to
different decoupled bands with band-indexes +, 0 and -
respectively. Right panel shows the low energy density of states
of AAA-stacked TLG normalized to its density of states at E=0,
$D(0)=\sqrt{2}g\gamma/4\pi v_{F}^{2}$.}\label{fig:01}
\end{center}
\end{figure}
\end{document}